# High Mobility Two-Dimensional Electron Gas at the BaSnO$_3$/SrNbO$_3$ Interface

Sharad Mahatara[1#], Suresh Thapa[2#], Hanjong Paik[3], Ryan Comes[2*] and Boris Kiefer[1*]

[1]Department of physics, New Mexico State University, Las Cruces, NM 88003-8001, USA
[2]Department of physics, Auburn University, Auburn, AL 36849, USA
[3]Materials Science and Engineering, Cornell University, Ithaca, NY 14850, USA

**ABSTRACT:** Oxide two-dimensional electron gases (2DEGs) promise high charge carrier concentrations and low-loss electronic transport in semiconductors such as BaSnO$_3$ (BSO). ACBN0 computations for BSO/SrNbO$_3$ (SNO) interfaces show Nb-*4d* electron injection into extended Sn-*5s* electronic states. The conduction band minimum consists of Sn-*5s* states ~1.2 eV below the Fermi level for intermediate thickness 6-unit cell BSO/6-unit cell SNO superlattices, corresponding to an electron density in BSO of ~$10^{21}$ cm$^{-3}$. Experimental studies of analogous SNO/BSO interfaces grown by molecular beam epitaxy confirm significant charge transfer from SNO to BSO. *In situ* angle-resolved X-ray photoelectron spectroscopy studies show an electron density of ~$4\times10^{21}$ cm$^{-3}$. The consistency of theory and experiment shows that BSO/SNO interfaces provide a novel materials platform for low loss electron transport in 2DEGs.

## INTRODUCTION

Complex oxide interfaces are well known for hosting a two-dimensional electron gas (2DEG) as a result of band engineering, a property which is absent in the corresponding bulk compounds.[1,2] An oxide 2DEG was first observed in LaAlO$_3$/SrTiO$_3$ (LAO/STO) heterostructures consisting of two wide band gap insulators LAO (E$_g$ = 5.5 eV, ref. 3) and STO (E$_g$ = 3.2 eV, ref. 4). The 2DEG has carrier density (~$10^{13}$ cm$^{-2}$),[5] strong confinement to the interface (~2 nm width),[6] and relatively high mobility at low temperature (~$10^5$ cm$^2$/Vs),[2] which makes it suitable for low-temperature electronic and optoelectronic applications.[7] Previously, defect engineering,[8] strain engineering,[9] and doping engineering[10] have been used to improve room temperature mobility. In parallel, novel heterostructures such as NdAlO$_3$/STO,[11] NdTiO$_3$/STO,[12] and LaTiO$_3$/STO[13] have been developed, but their room temperature charge carrier mobility sub-optimal due to the presence of localized Ti-*3d* orbitals in STO that are strongly scattered by longitudinal optical phonons.[14]

An alternative approach is to design oxide heterostructures with a conduction band minimum (CBM) that is dominated by less localized s-orbitals, thereby increasing band dispersion and carrier mobility.[15,16] BaSnO$_3$ (BSO) with a Sn-*5s* dominated CBM is an excellent candidate for the rational design of 2DEG heterostructures.[15] With a Hall mobility as high as 320 cm$^2$V$^{-1}$s$^{-1}$, BSO has the highest room temperature mobility in complex metal oxides to date with electron concentrations of $8\times10^{19}$ cm$^{-3}$ in a La-doped bulk BSO single crystal.[17] One of the synthesis challenges for BSO 2DEG formation is the prevention of band filling and compensating defects that occur at high dopant concentrations.[18–20] The second challenge is to select a suitable material for electron injection into BSO. Previous density-functional-theory (DFT) studies for metallic SrNbO$_3$ (SNO),[21,22] show Nb-*4d* $t_{2g}$ bands crossing the Fermi level and a large energy separation between O-*2p* and Nb-*4d* bands. These findings suggest possibly a large driving force for Nb-*4d* electron injection into BSO, and the realization of a 2DEG in the vicinity of BSO/SNO interfaces.

Here we use the synergy of self-consistent Hubbard-U DFT computations with experimental synthesis and spectroscopy to explore the electronic structure of BSO/SNO heterostructures. The consistency of the DFT results with our XPS experiments on MBE-grown heterostructures strongly suggests the presence of a high-mobility and high-density 2DEG at the BSO/SNO interface.

## METHODS

**Computations**. All computations were performed within the framework of 3D-periodic density-functional-theory (DFT) and complemented by a self-consistent Hubbard-U approach (ACBN0),[23–26] compatible with Quantum Espresso.[27,28] Following previous work,[29] ACBN0 Hubbard-U parameters were computed with norm-conserving pseudopotentials available in AFLOWπ,[24] and combined with the projector augmented wave (PAW) scheme[30] and pseudopotentials from the ps 1.0.0 library,[31] for all subsequent computations. Electronic exchange and correlation effects were described within the GGA as parametrized in PBE,[32] using a plane wave energy cutoff of E$_{cut}$ = 70 Ry. For cubic BSO and SNO crystal structures we used the DFT equilibrium structures to determine self-consistent ACBN0 Hubbard-U parameters (BaSnO$_3$: Ba = 3.96 eV, Sn = 0.04 eV, O = 8.88 eV; SrNbO$_3$: Sr = 0.05 eV, Nb = 1.07 eV, O = 7.39 eV), similar to previously reported Hubbard-U values for perovskite oxides.[33] With these Hubbard-U parameters we obtained an equilibrium lattice parameter for cubic BSO, a=4.118 Å (SNO: 4.044 Å), comparable to our DFT relaxed lattice parameter of 4.178 Å (SNO: 4.061 Å), and in excellent agreement with experiment, 4.115 Å[34] (SNO: 4.020 Å[35]). The BSO band gap increases from 0.4 eV (DFT) to 3.6 eV (ACBN0), and closer to of the experimental band gap of 3.1 eV,[36] while SNO remains non-magnetic metal, consistent with previous theory,[35] and experiment.[37] We built different (BSO)$N$/(SNO)$M$ heterostructures ($N$, $M$ refer to the number of cubic unit cells in each stack): (BSO)*4*/(SNO)*2*, (BSO)*4*/(SNO)*4*, (BSO)*6*/(SNO)*6*, (BSO)*7*/(SNO)*7* and (BSO)*8*/(SNO)*8*. The cross-sectional area and vertical dimensions of the heterostructures were fixed at the ACBN0 equilibrium lattice parameter of the BSO substrate, following previous work,[16,38,39] leading to an in-plane 1.8% tensile strain in the SNO layer. With these computational settings, we computed converged electronic structure, electronic density of states, effective mass at the CBM, and average charge density parallel to the [001] stacking directions (for further details see Supporting Information).

**Experiments.** The BSO/SNO heterostructure was created by growing SNO on as prepared BSO (001) single crystal films synthesized at Cornell University on Nb-doped STO (001) and GdScO$_3$ (110) substrates. SNO was synthesized at Auburn University using the hybrid molecular beam epitaxy (hMBE) approach described elsewhere using the tris(diethylamino)(t-butylimido)niobium (TBTDN) precursor.[22] SNO film thicknesses were between 3 and 5 unit cells in all cases to permit XPS measurements of the buried interface. Approximately 3 unit cells of SrHfO$_3$ (SHO) was deposited as a capping layer in some cases to preserve the surface during cooldown based on our previous observations of the surface stability of SNO.[22] *In situ* RHEED (Staib Instrument) was used to monitor the growth process and the quality of the film (for more details see Supporting Information). One challenge for films grown on Nb:STO is that strain relaxation in BSO produces a rougher initial surface compared to a bare substrate. Films grown on strained BSO below the critical thickness on GdScO$_3$ demonstrated smoother surfaces and a sharper RHEED pattern. For this work, we focus our analysis on three samples: a bare 10 nm BSO film on Nb-doped STO, an uncapped 4-unit cell SNO film on a 25 nm BSO film on Nb-doped STO, and an SHO-capped 4-unit-cell SNO film on a 3 nm BSO film on GdScO$_3$. After growth, samples were transferred from the MBE reactor to the PHI 5400 X-ray photoelectron spectroscopy (XPS) (Al Kα X-ray source, pressure ~1×10$^{-9}$ Torr) system through an ultra-high vacuum (UHV) transfer line. Angle-resolved XPS experiments were used to characterize the reduction of tin oxidation state from 4+ to 3+. An electron neutralizer gun was applied for compensating the charging of the insulating samples.[40] Analysis of the ARXPS data was performed using CasaXPS.[41,42] *Ex situ* atomic force microscopy measurements were carried out on an uncapped sample to check the morphology of SNO film (see Supporting Information).

## RESULTS AND DISCUSSION

The properties of (BSO)/(SNO) (001) interfaces were investigated by varying the number of cubic unit cells in the BSO (*N*) and/or SNO (*M*) layers and we considered: (BSO)4/(SNO)2, (BSO)4/(SNO)4, (BSO)6/(SNO)6, (BSO)7/(SNO)7 and (BSO)8/(SNO)8. For (BSO)6/(SNO)6 our computations predict the highest downward shift of the CBM relative to the Fermi level (~1.2 eV, Figure 1); and following ref. 18, we estimate a carrier concentration of ~10$^{21}$ cm$^{-3}$. All other heterostructures have smaller downward shifts, indicating lower charge carrier densities (Figure S1 and Table S1 in

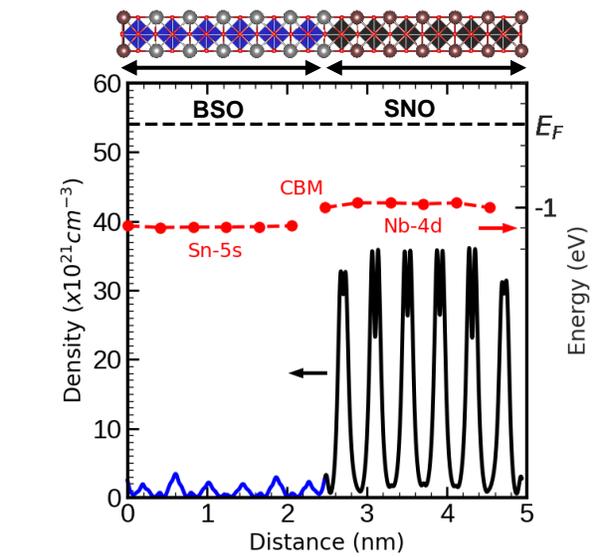

Figure 2. Planar average charge densities in (BSO)6/(SNO)6 plotted along [001] direction, blue line: BSO; black line: SNO. Layer-resolved CBM for (BSO)6/(SNO)6 are shown as filled red circles.

Supporting Information). Similarly, except for (BSO)8/(SNO)8, the CBM consists of Sn-*5s* electronic states, leading to an effective mass $m^*$ of 0.104 $m_0$ ($m_0$, electron rest mass), very similar to BSO bulk ($m^*$ = 0.098 $m_0$). Moreover, the effective mass varies by less than ~6% among these heterostructures (Table S1 in Supporting Information), and is ~4 times lower than that of the known 2DEG hosting LAO/STO (001) heterostructure, $m^*$= 0.4 $m_0$.[43] In contrast, for the (BSO)8/(SNO)8 heterostructure we find a ~65% increase of the effective mass (Figure S1 and Table S1 in Supporting Information), that is attributed to Nb-*4d* electronic states at the CBM. Therefore, our results confirm the hypothesized design principle of utilizing Sn-*5s* orbitals to increase mobile carrier concentration at BSO/SNO interfaces, facilitating 2DEG formation. The layer-resolved CBM for the (BSO)6/(SNO)6 heterostructure (Figure 2) shows that the CBM of the BSO layers (Sn-*5s*), is ~0.2 eV below the CBM of the SNO layers (Nb-*4d*), consistent with charge transfer from SNO to BSO.

The high electron charge carrier density of ~10$^{21}$ cm$^{-3}$ for the (BSO)6/(SNO)6 heterostructure is corroborated by the computed charge density variation perpendicular to the heterostructure stacking direction, [001] (Figure 2). The charge density in the BSO layer corresponds to the electron density of ~$10^{21}$ cm$^{-3}$ per unit cell, while electron densities are ~10x higher in the SNO layer. Moreover, the charge density in the SNO layer (1.3 x $10^{22}$ cm$^{-3}$ per unit cell) corresponds closely to the charge density inferred for complete depletion of Nb-$d^1$ states (1e$^-$ = 1.4 x $10^{22}$ cm$^{-3}$), supporting near-optimal charge transfer. Lastly, high electron accumulation at the (BSO)6/(SNO)6 interface is confirmed by the layer-resolved variation of the O-$2p$ center-of-mass (CM). The VBM in bulk SNO consists of O-$2p$ states and is located -5.2 eV below the Fermi energy (Figure S2 in Supporting Information), similar to previous ARPES experiments that reported -4.4 eV,[21] while the O-$2p$ CM is located ~3.0 eV below the VBM, -7.4 eV below the Fermi energy. In contrast, the O-$2p$ CM of BSO is located -2.6 eV below the Fermi energy, ~4.8 eV higher as compared to bulk SNO (Figure S2 in Supporting Information), generating the driving force for SNO to BSO charge transfer:[44] O-$2p$ CM alignment leads to a shifted Fermi level in SNO that is more positive than that of BSO. The thermodynamic requirement of a well-defined and unique Fermi level throughout the heterostructure is restored through charge transfer from Nb-$d^1$ to Sn-$5s$ orbitals, in agreement with our computations (Figure 3). Therefore, due to the charge transfer from SNO to BSO across the interface, charge depletion in near-interface SNO layers is expected, in excellent agreement with the reduced real space charge density amplitude near the BSO/SNO interface (Figure 2).

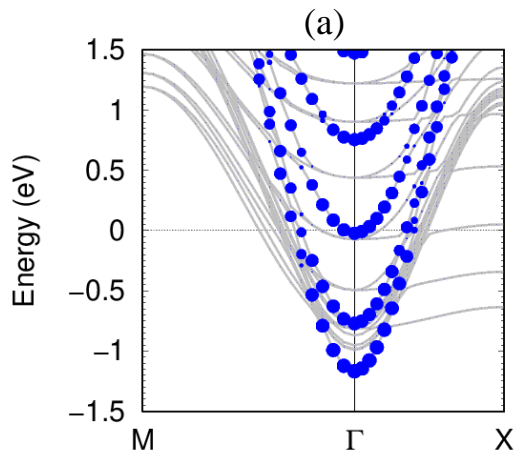
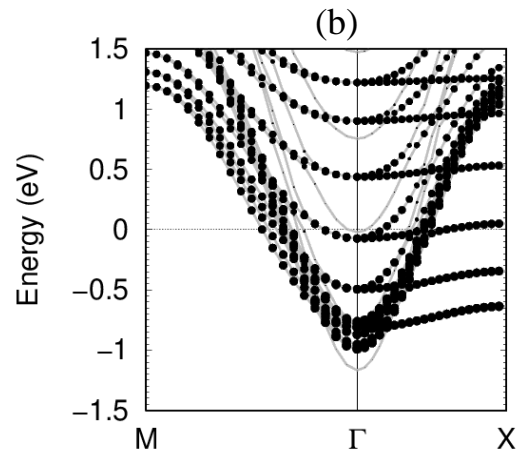

Figure 1. Orbital projected bandstructure of (BSO)6/(SNO)6 heterostructure. The blue and black circles correspond to Sn-*5s* and Nb-*4d*, respectively.

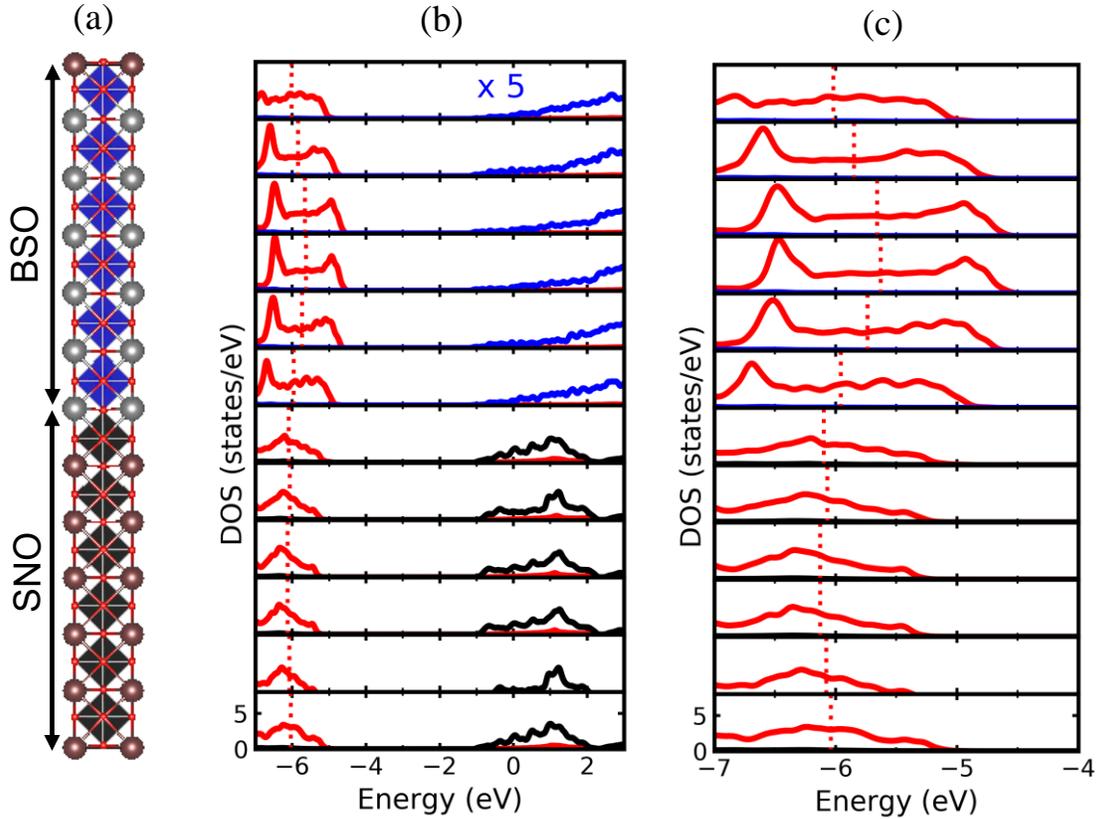

Figure 3. (BSO)6/(SNO)6 heterostructure: (a) atomic structure, blue and black octahedra correspond to Sn and Nb coordination, respectively. (b) layer-projected partial density of states (pdos) of O-$2p$ (red), Sn-$5s$ (blue) and Nb-$4d$ (black). The pdos Sn-$5s$ (blue) is zoomed in by 5 times, (c) pdos of O-$2p$ taken between 7 to 4 eV below the Fermi energy. The vertical dotted line (red) is the center-of-mass of O-$2p$, relative to the Fermi energy at E=0 eV.

To experimentally study the DFT predicted charge transfer, and Sn$^{4+}$ to Sn$^{3+}$ reduction, BSO/SNO heterostructures were synthesized and characterized as described above. The two heterostructures are represented schematically in Figure 4(a). The Sn-$3d$ core level spectra for an uncapped sample and a SHO capped sample in reference to the plasma cleaned BSO on Nb:STO substrate at 45° and 70° photoelectron emission angles are shown in Figure 4(b). In an uncapped sample, an angle-resolved XPS (ARXPS) measurements of the Sn-$3d$ core level at shallower angle shows a more prominent additional shoulder at lower binding energy compared to the 45° orientation. This extra shoulder is assigned to Sn$^{3+}$ associated with charge-transfer-induced Sn$^{4+}$ reduction. Additionally, Sn-$3d_{5/2}$ core level deconvolution of a substrate along with SHO-capped sample grown on is carried out using the same constraints as for the uncapped sample. A plasma-cleaned bare BSO on Nb:STO substrate shows no secondary peak at lower binding energy. To quantify this lower oxidation state associated with Sn, the Sn-$3d_{5/2}$ core level of the uncapped sample is deconvoluted by constraining a second component with the same FWHM as the Sn$^{4+}$ component and at 1.75 eV lower binding energy than the Sn$^{4+}$ component as shown in Figure 4b for 70° photoelectron emission angle.

Deconvolution of a capped sample demonstrates that the SHO capping has increased the area of the Sn$^{3+}$ shoulder at lower binding energy compared to the uncapped sample at 45° emission. This result suggests that the capping of the SNO film surface is a key ingredient for SNO to donate mobile charge carriers to the BSO interface layer. These free electrons fill Sn-*5s* states as reported previously,[11,45] improves charge carrier mobility, and facilitates 2DEG formation. For the purpose of estimating charge transfer, however, we choose to focus on the uncapped sample for better depth resolution into the BSO layer.

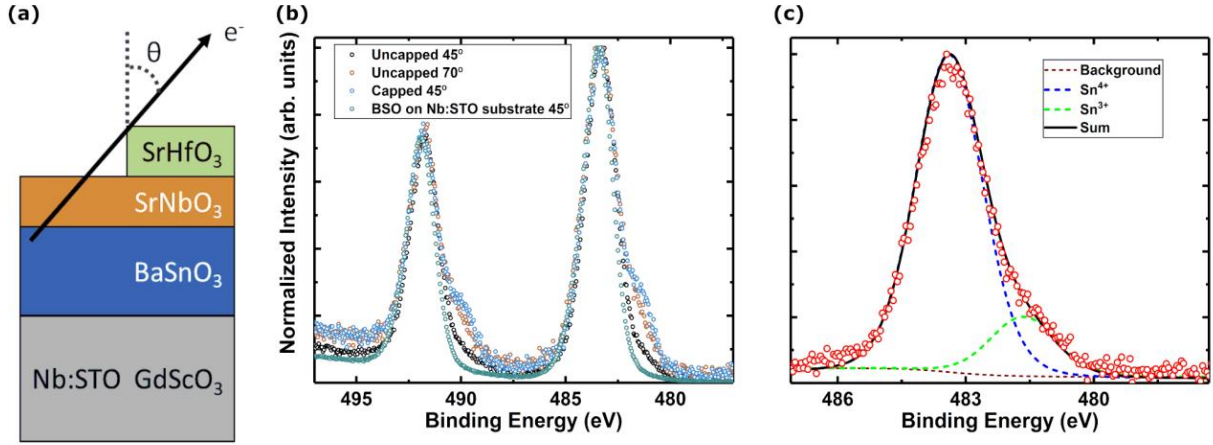

Figure 4. (a) Schematic of heterostructures studied by angle-resolved XPS including emission angle θ, showing uncapped sample on Nb:STO and capped sample on GdScO$_3$; (b) Sn-*3d* core level for uncapped sample at θ=45° and θ=70° emission angles and SHO-capped sample substrate at θ=45°, and (c) Sn-*3d$_{5/2}$* core level of an uncapped sample deconvolved for 70° photoelectron emission angle.

The Sn$^{3+}$ abundance in the uncapped sample was obtained by deconvoluting the Sn-*3d$_{5/2}$* core level as a function of photoelectron emission angle collected from 45° to 70° at an interval of 5° in ARXPS analysis (Figure 5). As the photoelectron emission angle becomes shallower, an increasing trend of Sn$^{3+}$ concentration in the BSO layer is observed. This result shows that the Sn$^{3+}$ charge state is greater towards the surface of the heterostructure, consistent with an accumulation of electrons at the BSO/SNO interface. Such measurements have also been employed in the past for a variety of oxide heterostructures to examine cation intermixing and electron accumulation at an interface from charge transfer.[46] More sophisticated modeling of the electron concentration within each unit cell of the BSO film below the interface can also be used to estimate the electron density within the BSO layer.

By assuming a reduction in the electron concentration of 33% per BSO unit cell moving away from the interface, which mimics the DFT results above, we can derive a model to estimate the Sn$^{3+}$ distribution in BSO (Figure 5b). Due to the high degree of Sn-*5s* states in BSO, a significantly greater electron diffusion into the film would be expected in comparison to the more localized transition metal d-states that are commonly studied in oxide interfaces. We assume an inelastic mean free path (IMFP) of 20 Å, suitable for the kinetic energy of Sn-*3d* electrons passing through the SNO layer and a BSO layer thickness of 40 Å (9-unit cells). This model allows us to predict the contributions of electrons at various depths within BSO to the ARXPS experimental data. An analogous model was employed previously to estimate cation intermixing in LaCrO$_3$/SrTiO$_3$ superlattices.[47] This analysis shows that a model with ~35-40% Sn$^{3+}$ concentration at the BSO/SNO interface fits well with the ARXPS results. However, lab-based ARXPS is relatively insensitive to electron diffusion into BSO, which merits further study using advanced techniques such as hard X-ray photoemission (HAXPES).[48,49] We emphasize that these results rely on several assumptions and are not a unique "best-fit" to the ARXPS data, but provide an estimate of the charge transfer from SNO to BSO that accounts for electron diffusion into the BSO layer.

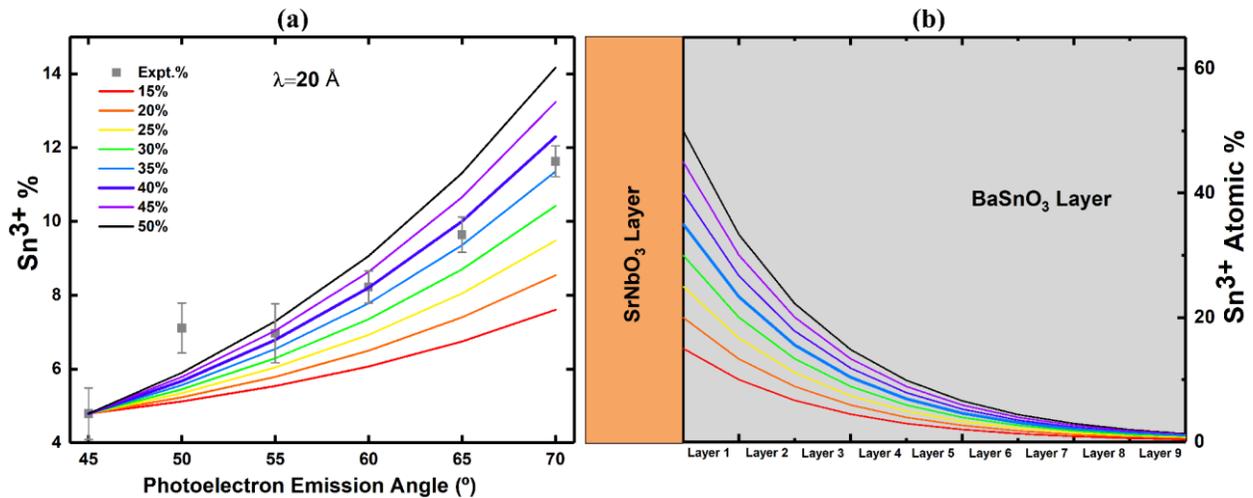

Figure 5. (a) Experimental Sn$^{3+}$ peak area ratio as a function of photoelectron emission angle scatter plot and model fitted for different concentrations and (b) Sn$^{3+}$ atomic percentage decay model as a function of BSO thickness.

Taking 4.1 Å as a lattice parameter for the BSO cubic structure, we have estimated the volume electron density and sheet carrier concentration at the interface BSO layer for 35% $Sn^{3+}$ electronic states. The estimated electron density of $4\times10^{21}$ cm$^{-3}$ in the interfacial BSO layer of BSO/SNO interface is significantly higher than other reports for La-doped BSO films,[15,16] La-doped BSO single crystals,[17,50] and oxygen-deficient BSO films[51] by at least a factor of 4 in all cases. The sheet electron concentration estimated from the ACBN0 model is ~$0.4\times10^{14}$ cm$^{-2}$ at the interface, which also agrees semi-quantitatively with the sheet electron concentration estimated from the ARXPS model (>$1.8\times10^{14}$ cm$^{-2}$) for only the interfacial layer of the BSO film. This charge carrier value as estimated from the ARXPS model is roughly an order of magnitude larger than those reported for LaInO$_3$/BSO polar/non-polar interfaces[45,52,53] and ~30x larger than modulation-doped La:SrSnO$_3$/BSO heterostructures.[48]

Differences between theory and experiment are attributed to the larger BSO bandgap (3.6 eV) in the ACBN0 computations, in comparison to the experimental band gap of ~3.1 eV.[54] This difference would lead to an underestimate of the charge transferred from SNO to BSO. Additional variation could occur due to out-diffusion of Sn cations into the SNO that would complicate the ARXPS model. Finally, while we cannot rule out the presence of some growth-induced oxygen vacancies at the interface, we do not believe that our results can be explained solely based on oxygen vacancies (for details see Supporting Information).

Future studies in this area should focus on determination of transport behavior and carrier dynamics in these heterostructures using models that account for the parallel conduction pathways within the BSO 2DEG and the partially electron-depleted SNO layer. The models developed for charge transfer provide an estimate of the expected sheet carrier concentration within BSO, but decoupling the contributions of BSO and SNO to transport behavior is likely to prove challenging. Efforts have been made to decouple the transport contributions from distinct layers in similar La:SSO/BSO,[48] SrTaO$_3$/STO,[55] and SNO/STO heterostructures,[56] but care must be taken to account for the surface instability of SNO in these models. We suggest that THz spectroscopy studies may provide a promising avenue for further examination of these 2D electronic systems.[57] Future studies could also explore the growth of SNO on thicker BSO layers such as single-crystal BSO[50] or on films grown on a lattice matched substrate such as Ba$_2$ScNbO$_6$,[58] which would open up easier pathways for device integration.

## CONCLUSIONS

In conclusion, Hubbard-U augmented DFT computations show that BSO/SNO superlattices exhibit charge transfer and high mobility for 2DEG formation in the BSO layer. For thin heterostructures the conduction band minimum (CBM) is dominated by Sn-$5s$ bands and located well below the Fermi energy. The CBM corresponds to the low effective mass of ~0.10 m$_0$ and varies by less than ~6% for thin heterostructures, and remains very similar to high-mobility bulk BSO. The CBM in (BSO)$6$/(SNO)$6$ heterostructure shows the highest CBM suppression below the Fermi energy, ~1.2 eV, corresponding to a high electron density of ~$10^{21}$ cm$^{-3}$. Thinner heterostructures show less pronounced CBM suppression. In contrast, the CBM in thicker heterostructures is dominated by Nb-$4d$ electronic states. BSO/SNO heterostructures grown by hMBE confirm the expected charge transfer, based on *in situ* XPS analysis of the Sn-$3d$ core level peak. The quantitative analysis of the ARXPS results shows that the capping layer of SHO enhances charge transfer across the BSO/SNO interface. The interfacial BSO electron density from ARXPS is ~$4\times10^{21}$ cm$^{-3}$, in semi-quantitative agreement with our ACBN0 predicted electron density of ~$10^{21}$ cm$^{-3}$. Integration across the depth of the film produces a total charge density in the BSO layer of ~$10^{14}$ cm$^{-2}$. Therefore, BSO/SNO heterostructures are promising materials for hosting a high mobility 2DEG and the study of their emergent properties.

## AUTHOR INFORMATION

*Corresponding authors.

#Authors contributed equally to this work.

## ACKNOWLEDGMENT

The authors thank Marcelo Kuroda of Auburn University for helpful discussions. BK and SM would like to acknowledge computing support by the Extreme Science and Engineering Discovery Environment (XSEDE) resource Stampede2 at TACC through allocation TG-DMR110093. ST and RBC gratefully acknowledge support from the Air Force Office of Scientific Research under award number FA9550-20-1-0034 and Alabama EPSCoR-GRSP Fellowship. HP acknowledges partial support through the National Science Foundation (Platform for the Accelerated Realization, Analysis, and Discovery of Interface Materials (PARADIM)) under Cooperative Agreement No. DMR-2039380.

# Supporting Information

# High Mobility Two-Dimensional Electron Gas at the BaSnO$_3$/SrNbO$_3$ Interface

Sharad Mahatara[1#], Suresh Thapa[2#], Hanjong Paik[3], Ryan Comes[2*] and Boris Kiefer[1*]

[1]Department of physics, New Mexico State University, Las Cruces, NM 88003-8001, USA
[2]Department of physics, Auburn University, Auburn, AL 36849, USA
[3]Materials Science and Engineering, Cornell University, Ithaca, NY 14850, USA

**Computational Details**

Convergence tests showed that increasing the cutoff from 70 Ry to 80 Ry changes the total energy by less than 1.0 meV/atom. The k-points grid of size 12x12x12 and 12x12x1 were used for the bulk and interface calculations, respectively. Increasing the k-points grids to 14x14x14 and 14x14x1 for the bulk and interface changes the total energy by less than 1.0 meV/atom. The charge density for the bandstructure computations was computed with a k-point grid of size 12x12x12 (bulk) and 12x12x1 (interface), while all orbital and site-projected quantities were computed with denser k-point grids of 30x30x30 and 30x30x1 for bulk and interface calculations, respectively. We computed the electron densities by integrating layer-resolved partial densities of states of conduction bands below the Fermi energy. The effective mass at the CBM was obtained from polynomial fits of the electronic dispersion in the vicinity of the CBM, following previous work.[1]

**Experimental Methods**

Changes implemented are the growth chamber shroud walls were maintained at -60 °C compared to -30 °C via a closed loop chiller and low temperature fluid (Syltherm XLT, Dow Chemical) to effectively reduce the background water vapor pressure from the dissociated TDTBN molecules and the hafnium precursor, tetrakis(ethylmethylamino) hafnium (TEMAH) is used for growth of the SHO capping layer. Hafnium was supplied through a gas source using a metalorganic precursor, (TEMAH) (99.99%, Sigma-Aldrich, USA) from a bubbler connected to the growth chamber without carrier gas.

**DFT Results and Discussion**

The computed band structures confirm the expectation that the CBM is dominated by Sn-*5s* electronic states for (BSO)*4*/(SNO)*2*,

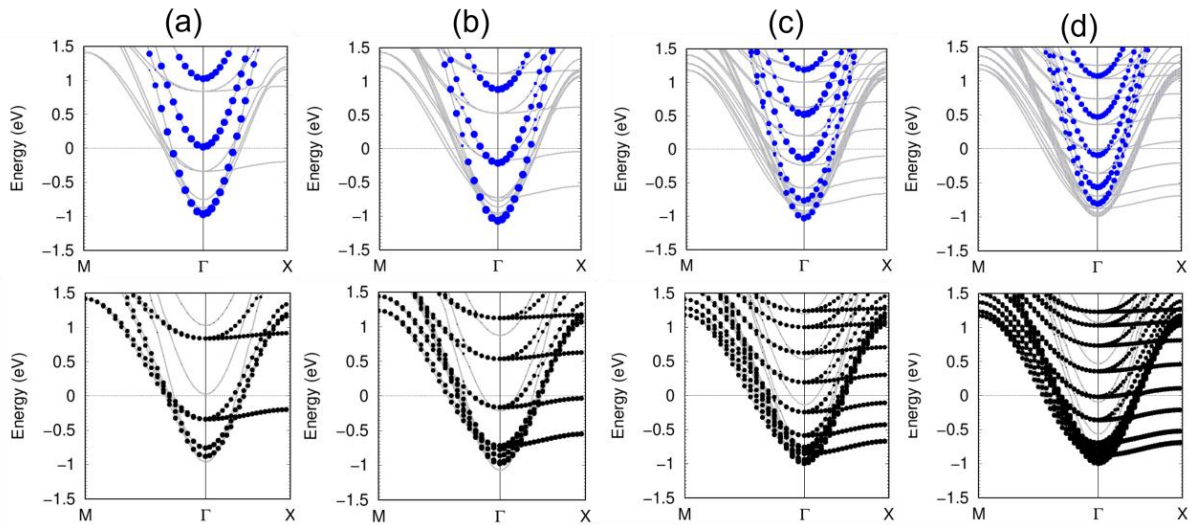

Figure S1. Orbital projected bandstructure of (BSO)*N*/(SNO)*M* heterostructures (*N, M* refer to the number of cubic unit cells in each stack): (a) (BSO)*4*/(SNO)*2*, (b) (BSO)*4*/(SNO)*4*, (c) (BSO)*7*/(SNO)*7*, and (d) (BSO)*8*/(SNO)*8* heterostructure. The blue and black bands correspond to Sn-*5s* and Nb-*4d*, respectively.

(BSO)*4*/(SNO)*4,* (BSO)*6*/(SNO)*6* and (BSO)*7*/(SNO)*7* heterostructures. In contrast, (BSO)8/(SNO)8 shows that the CBM is dominated by Nb-*4d* electronic states (Figure S1). Among the tested BSO/SNO heterostructures, (BSO)*6*/(SNO)*6* is most promising since the Sn-*5s* states are located furthest below the Nb-*4d* states, reducing inter-band electronic transitions and electronic scattering. Moreover, the downward shift in the (BSO)*6*/(SNO)*6* heterostructure is highest (~1.2 eV, see main text, Figure 1) while the downward shifts for the other heterostructures are 0.98 eV, 1.08 eV, 0.8 eV and 0.98 eV for (BSO)*4*/(SNO)*2*, (BSO)*4*/(SNO)*4*, (BSO)*7*/(SNO)*7,* and (BSO)*8*/(SNO)*8* heterostructures, respectively, whcih following Ref. 2 and 3 corresponds to a lower charge carrier density (Table S1). Therefore, (BSO)6/(SNO)6 heterostructure is expected to show the highest volume charge density (Table S1).

Table S1. Volume charge density and effective mass at the CBM for the investigated heterostructures.

| | (BSO)*4*/(SNO)*2* | (BSO)*4*/(SNO)*4* | (BSO)*6*/(SNO)*6* | (BSO)*7*/(SNO)*7* | (BSO)*8*/(SNO)*8* |
|---|---|---|---|---|---|
| Charge density per unit cell (x$10^{21}$ cm$^{-3}$) | 0.75 | 0.94 | 0.99 | 0.83 | 0.73 |
| CBM Effective mass ($m^*/m_0$) | 0.102 | 0.102 | 0.104 | 0.108 | 0.168 |

The effective mass of the CBM up to and including the (BSO)7/(SNO)7 heterostructure is ~4% higher than the computed effective mass of bulk BSO (0.098 $m_0$) and varies weakly with layer thickness as long as the CBM consists of Sn-*5s* bands (Figure 1, main text, Figure S1 and Table S1). The CBM effective mass is ~4 times lower than LAO/STO (001).[4] Additionally, the bands become progressively heavier at higher energy due to the dominant contribution of Nb-*4d* bands (Figure S1). The electron mobility at the BSO/SNO interface is expected to be higher than that of LAO/STO for two reasons: first, the effective mass of 2DEG at BSO/SNO heterostructure (~0.10 $m_0$) is ~4 times lower than effective mass of the lightest $d_{xy}$ electrons at the LAO/STO (001) interface (~0.4 $m_0$);[4] second, Sn-*5s* states are furthest below the Nb-*4d* states for (BSO)6/(SNO)6 heterostructure, reducing inter-band electronic transitions and electronic scattering.

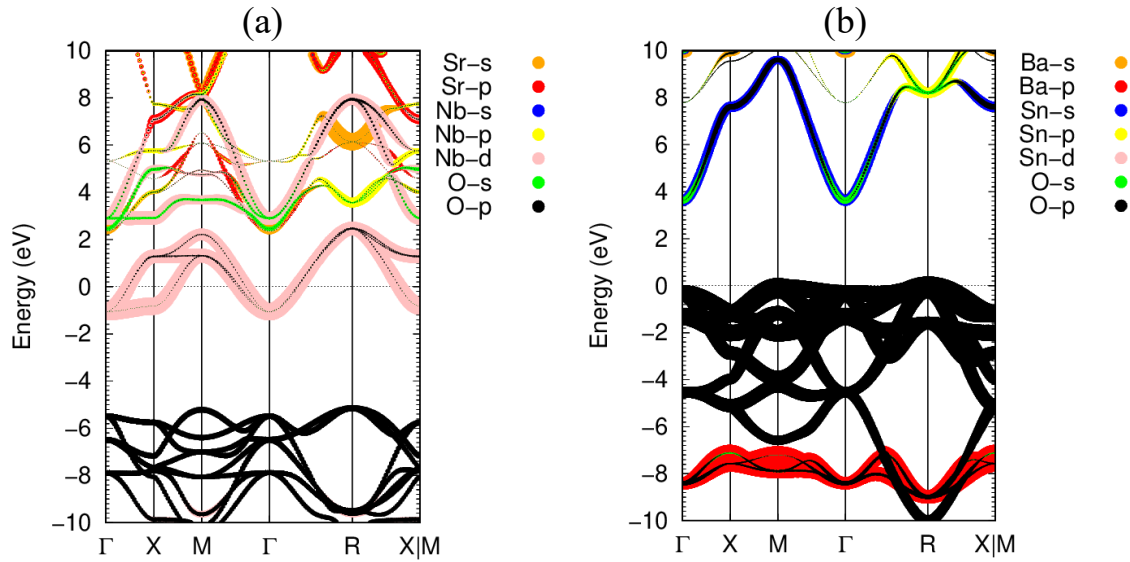

Figure S2. Orbital projected bandstructure of the bulk (a) SNO and (b) BSO.

The computed ACBN0 bandstructure confirms that bulk SNO is metallic, with Nb-*4d* states dominating the electronic structure at the Fermi energy (Figure S2a). The fully occupied valence band is dominated mainly by O-*2p*. The VBM in SNO is located -5.2 eV below the Fermi energy, similar to previous ARPES experiment that reported -4.4 eV.[5] The O-*2p* band extends between -5.2 eV and -10.5 eV, with center-of-mass (CM) at -7.4 eV. The ACBN0 computed bandgap for bulk BSO is ~3.6 eV (Figure S2b), significantly larger than the DFT-only band gap of ~0.4 eV, and in much better agreement with the experimental value of 3.1 eV.[6] The CBM in BSO is dominated by Sn-*5s* orbitals, while the VBM is dominated by O-*2p*, similar to the bulk SNO. The O-*2p* dominated valence band extends between 0 eV and -6.6 eV, with a CM at -2.6 eV, 4.8 eV more positive than the SNO O-*2p* CM, in support of significant charge transfer from SNO to BSO, as stated in the main text.

**Experimental Results and Discussions**

Our results for the electron concentrations calculated from XPS cannot be explained solely on the basis of oxygen vacancies. The samples were heated in oxygen plasma prior to growth to prevent reduction of the BSO and the plasma was only turned off at the beginning of the growth process, limiting the time for the BSO to reduce. Additionally, the intensity of the $Sn^{3+}$ peak increases when the film is capped with $SrHfO_3$, which suggests that it is more sensitive to the number of electrons in SNO available for charge transfer rather than the oxygen environment during growth. Finally, predictions of BSO carrier concentrations in oxygen poor environments by DFT have indicated that carrier concentrations of ~$10^{18}$-$10^{19}$ cm$^{-3}$ would be expected.[7] Previous work has shown that oxygen-deficient BSO films have electron concentrations of only ~$10^{19}$ cm$^{-3}$,[8] which is well below the values obtained for our samples. Studies that incorporate BSO growth immediately followed by SNO growth would be valuable for further analysis in this regard, but are not possible in our MBE system.